\documentclass{PoS}

\title{Hercules X-1: variations of the cyclotron line energy with flux, with phase and with time}

\ShortTitle{Hercules X-1}

\author{\speaker{R\"udiger Staubert}\thanks{with D. Klochkov, D. Vasco, K. Postnov, 
N. Shakura, J. Wilms, R. Rothschild, A. Santangelo, B. Grefenstette, F. F\"urst, F. Harrison and others}\\
        Institut f\"ur Astronomie und Astrophysik, Universit\"at T\"ubingen,
        Sand 1, D-72076 T\"ubingen, Germany\\
        E-mail: \email{staubert@astro.uni-tuebingen.de}}

\abstract{Her X-1 is one of the the most intensively studied Accreting X-ray Binary Pulsars (AXBPs). 
This is largely because it is a bright and persistent X-ray pulsar, showing the largest variety of 
observable phenomena, partly due to the favorable geometry in observing the binary nearly edge-on. 
Her X-1 is the binary X-ray pulsar in which the first cyclotron line (or Cyclotron Resonant Scattering 
Feature - CRSF) was detected in 1976. Since then, we have made an effort to observe the source 
and the CRSF around 40\,keV as often as possible, using those X-ray satellites which cover the high 
energy X-ray range. Extended contributions are from the \textsl{Rossi X-ray Timing Explorer} (RXTE),
\textsl{INTEGRAL}, \textsl{Suzaku} and \textsl{Swift}. The most recent contribution with a new degree
of quality is from the high energy imaging telescope on \textsl{NuStar}. We have found that the CRSF
is variable with flux, with phase (both pulsational and precessional phase) and with time. 
The historical evolution of the pulse phase averaged CRSF centroid energy E$_{\rm cyc}$ since its discovery
is characterized by an initial value around 35\,keV, an abrupt jump upwards to beyond $\sim40$\,keV
between 1990 and 1994 and an apparent secular decay thereafter. Much of this decay, 
however, was found to be due to an artifact, namely a correlation between E$_{\rm cyc}$ and the X-ray 
luminosity $L_{x}$ discovered in 2007, amounting to a change of $\sim$ 7\% in energy for a factor of 
two in luminosity. In observations since 2007, however, we now find a statistically significant trend of
a true secular decrease of the cyclotron line energy. In addition, we discuss recent results of
pulse phase spectroscopy and evidence for the suspected variation in E$_{\rm cyc}$ with phase of 
the 35\,d precessional period (which is observed both in the modulation of the X-ray flux and in the systematic 
variation in shape of the 1.24\,s pulse profile).
}

\FullConference{"An INTEGRAL view of the high-energy sky (the first 10 years)"
9th INTEGRAL Workshop and celebration of the 10th anniversary of the launch,\\
		October 15-19, 2012\\
		Bibliotheque Nationale de France, Paris, France}

\begin{document}

%Table 1 -------------------------------------------------------------------
\begin{table}
\caption[]{Details of recent observations of Her~X-1 by \textsl{INTEGRAL}, \textsl{RXTE} 
\textsl{Suzaku} and \textsl{NuStar}.}
%\indent~~~~~~~~~~~~~~~as proposed by our group.}
    \label{observations}
\vspace{-3mm}
\begin{center}
\begin{tabular}{lllll}
%\begin{tabular}{lllll}
\hline\noalign{\smallskip}
\vspace{-1mm}
Observatory          & Date of               & Center       & Obs ID                      & Exposure  \\
                              & observation        & MJD           &                                  & [ksec]    \\
\hline\noalign{\smallskip}
\textsl{INTEGRAL}& 2007 Sep 03-08  & 54348.86  &  Rev. 597/598           & 414.72  \\   %Staubert et al. \\
\textsl{RXTE}        & 2009 Feb 04/05   & 54866.95  &  P 80015                   & ~~22.19   \\   %Staubert et al. \\ 
\textsl{Suzaku}      & 2010 Sep 22       & 55461.63  &  405058010/20          &  ~~41.66   \\   %Staubert et al. \\
\textsl{Suzaku}      & 2010 Sep 29       & 55468.51  &  405058030/40          &  ~~45.66   \\   %Staubert et al. \\
\textsl{INTEGRAL}& 2010 July 10-18  & 55391.43  &  Rev. 945-947            & 621.03  \\   %Shakura et al. \\
\textsl{Suzaku}      & 2012 Sep 19-25  & 56192.23  &  4070510-10,20,30    &  $\sim 70$   \\   %Staubert et al. \\
\textsl{NuStar}       & 2012 Sep 19-25  & 56192.23  &  3000200600-2,3,5,7 &  ~~72.9   \\   %Staubert et al. \\
\noalign{\smallskip}\hline
\end{tabular}\end{center}
\end{table}
%Table 1 -------------------------------------------------------------------

\section{Introduction}

\vspace{-3mm}
 The X-ray spectrum of the accreting binary pulsar Her~X-1 is
characterized by a power law continuum with exponential cut-off and an
apparent line-like feature, which was discovered in 1975 (Tr\"umper et al. 1978).
This feature is now generally accepted as an absorption feature around
40\,{\rm keV} due to resonant scattering of photons off electrons on
quantized energy levels (Landau levels) in the Teragauss magnetic
field at the polar cap of the neutron star. The feature is therefore
often referred to as a Cyclotron Resonant Scattering Feature (CRSF).  The
energy spacing between the Landau levels is given by E$_{\rm cyc}$ =
$\hbar eB/m_{\rm e}c$ = $11.6\,{\rm keV}\,B_{12}$, where
$B_{12}=B/10^{12}\,\rm{G}$, providing a direct method of measuring
the magnetic field strength at the site of the generation of the X-ray
spectrum. The observed line energy is subject to gravitational
redshift, $z$, such that the magnetic field may be estimated by
$B_{12}$ = (1+z)~$E_{\rm obs}/11.6\,{\rm keV}$.  The discovery of the
cyclotron feature in the spectrum of Her X-1 provided the first ever
`direct measurement' of the magnetic field strength of a neutron star,
in the sense that no other model assumptions are needed.  Originally
considered an exception, cyclotron features are now known to be rather
common in accreting X-ray pulsars, with more than a dozen binary
pulsars being confirmed cyclotron line sources, with several objects
showing multiple lines (up to four harmonics in 4U~0115+63). Reviews are 
given by e.g. Coburn et al. 2002, Staubert 2003, Heindl et al. 2004,
Terada et al. 2007, Wilms 2012.

\indent Here we summarize some results, including our most recent ones, about 
the variability in cyclotron line energy.  The application of pulse phase spectroscopy 
techniques has demonstrated
% (with  unprecedented statistical accuracy) 
a very strong 
variation in  E$_{\rm cyc}$ as function of pulse phase. Considering this pulse phase 
variation and the variation in pulse shape with 35\,d phase, there must be a 
variation in the phase averaged E$_{\rm cyc}$ with 35\,d phase. 
%We compare the expected variation with first observational results.
From a long-term study, we present the first statistically significant evidence for a 
true secular decrease of the phase averaged E$_{\rm cyc}$. We speculate about
the physics behind the secular decrease as being connected to changes in the 
configuration of the magnetic field structure at the polar caps of the accreting neutron star.

%Table 2 -------------------------------------------------------------------
\begin{table}
\caption[]{Recent cyclotron line energy measurements in Her~X-1 by
\textsl{INTEGRAL}, \textsl{RXTE}, \textsl{Suzaku} and \textsl{NuStar}. 
%from observations proposed by our group. 
Uncertainties are at the 68\% level. The two 
\textsl{Suzaku} data points from 2005 and 2006 are from Enoto et al. (2008),
adjusted to describing the cyclotron line by a Gaussian line profile (see text).
The \textsl{NuStar} value, as provided by the \textsl{NuStar} team, is preliminary.
%The 2005 data points are reproduced from Table~1 of \citet{staubert_etal07}.
%from \textsl{INTEGRAL} is from \citet{Klochkov_etal06}. 
35\,d cycle numbering is according to Staubert et al. (1983, 2009a).}
       \label{results}
\begin{center}
\vspace{-2mm}
\begin{tabular}{llllll}
%\begin{tabular}{lllll}
\hline\noalign{\smallskip}
\vspace{-1mm}
Satellite & Observation & 35\,d & Center    & Cyclotron Line         & max. Flux \\
              & month/year  & cycle & [MJD]        & Energy [keV]         & [ASM cts/s]$^{*}$ \\
\hline\noalign{\smallskip}
%\textsl{RXTE} \\       
%July 96     & 257   & 50029.75  & $41.12\pm0.55$  & $7.37\pm0.34$ \\
%Sep 97      & 269   & 50707.06  & $40.62\pm0.49$  & $7.49\pm0.73$ \\
%Dec 00      & 304   & 51897.69  & $40.07\pm0.31$  & $6.04\pm0.47$ \\
%Jan 01      & 305   & 51933.67  & $39.05\pm0.55$  & $5.72\pm0.34$ \\
%May 01      & 308   & 52035.48  & $39.93\pm0.63$  & $7.15\pm0.50$ \\
%June 01     & 309   & 52071.16  & $39.73\pm0.52$  & $6.93\pm0.20$ \\
%Dec 01      & 314   & 52245.09  & $40.04\pm0.22$  & \\
%Aug 02      & 321   & 52492.96  & $40.01\pm0.29$  & $7.19\pm0.26$ \\
%Nov 02      & 324   & 52599.32  & $40.51\pm0.13$  & $7.64\pm0.30$ \\
%Dec 02      & 325   & 52634.01  & $40.60\pm0.41$  & $7.55\pm0.34$ \\
%Oct 04      & 344   & 53300.95  & $38.51\pm0.51$  & $4.50\pm0.24$ \\
%July 05     & 352   & 53577.35  & $38.95\pm0.52$  & $5.12\pm0.37$ \\
\textbf{\textsl{RXTE}} & Feb 2009       & 388 & 54866.60    &  $37.58\pm0.65$   &  $5.94\pm0.50$ \\ %(Staubert et al. observation)   
\textbf{\textsl{Suzaku}} & May 2005       & 353  &  53648.90  &  $38.70\pm1.00$   &  $4.80\pm0.35$ \\ %(Enoto et al., corrected)
                                     & Mar 2006        & 358  &  53823.90  &  $38.10\pm1.00$   &  $4.70\pm0.35$ \\ %(Enoto et al., corrected)
                                     & Sep 2010        & 405  &  55461.60  &  $37.80\pm0.34$   &  $7.56\pm0.23$ \\ %(Staubert et al. observation)
                                     & Sep 2012        & 426  &  56192.23  &  not yet analysed   &  $6.94\pm0.20$$^{*}$   \\ %(Grefenstette et al. obs)
\textbf{\textsl{INTEGRAL}} & July/Aug 2007 & 373 & 54348.00   &  $36.50\pm0.60$   & $4.50\pm0.48$ \\ %(Staubert et al. observation)
                                           & July 2010        & 403 & 55389.30   &  $38.00\pm0.60$   & $8.50\pm0.24$ \\ %(Shakura et al. observation) 
                                           & June 2011       & 413 & 55738.47   &  $37.34\pm0.28$   & $7.00\pm0.20$ \\ %(Staubert et al. observation, Main-On)
%July 2012      & 421 & 55744.40   &  \\ %(Staubert et al. observation, Main-On)
%July 2011      & 421 & 55759.36   &  \\ %(Staubert et al. observation, Short-On)
%July 2011 \\ %(Staubert et al. observation)
\textbf{\textsl{NuStar}} & Sep 2012         & 426  &  56192.23  &  $37.30\pm0.15$  & $6.94\pm0.20$$^{*}$  \\ %(Grefenstette et al. obs)
\noalign{\smallskip}\hline
\end{tabular}
\end{center}
\vspace{-1mm}
$^{*}$ The maximum \textsl{Main-On} flux was determined using the monitoring data of \textsl{RXTE}/ASM
and \textsl{Swift}/BAT (since 2012 from BAT only); the conversion is: (2-10\,keV ASM cts/s) = 90 * 
(15-50\,keV BAT cts~cm$^{-2}$~s$^{-1}$)
\end{table}
%Table 2 -------------------------------------------------------------------

\section{Data base and method of analysis}
\vspace{-3mm}
Her~X-1 is probably the best observed accreting binary X-ray pulsar.
Its X-ray spectrum and the cyclotron line (CRSF) has been measured by many
instruments since the discovery of the CRSF in 1975 (Tr\"umper et al. 1978). 
The data base behind previously reported results are summarized in
corresponding Tables of the following publications: Gruber et al. 2001, Coburn et al. 2002,
Staubert et al. (2007, 2009a, 2013), Klochkov et al. (2008, 2011) and Vasco et al. (2011, 2013). 
Details about more recent observations (proposed by our group during the last five years) 
by \textsl{RXTE}, \textsl{INTEGRAL}, \textsl{Suzaku} and \textsl{NuStar}
%, of which results are included here, 
are given in Tables \ref{observations} and \ref{results}.
Generally, Main-On state observations (35\,day-phases $<0.22$)
%, where the X-ray flux is the highest) 
were used. The spectral analysis was performed using 
the standard software appropriate for the respective satellites (see the 
publications cited above). Unless otherwise stated, the spectral model used was
a power law with exponential cut-off (\texttt{highecut, cutoffpl, FermiDirac}) for
the continuum and a multiplicative absorption line with a Gaussian optical depth 
profile for the cyclotron line. Details of the fitting procedure can be found in 
the papers cited above. Here we discuss results of the spectral analyses of pulse 
phase resolved and pulse phase averaged spectra, a dependence on 35\,phase
and on elapsed time.

%\newpage

%\vspace{-40mm}
% Fig. 1  new ------------------------------------------------------------------------------------------------------------
\begin{figure}
\includegraphics[width=7.7cm]{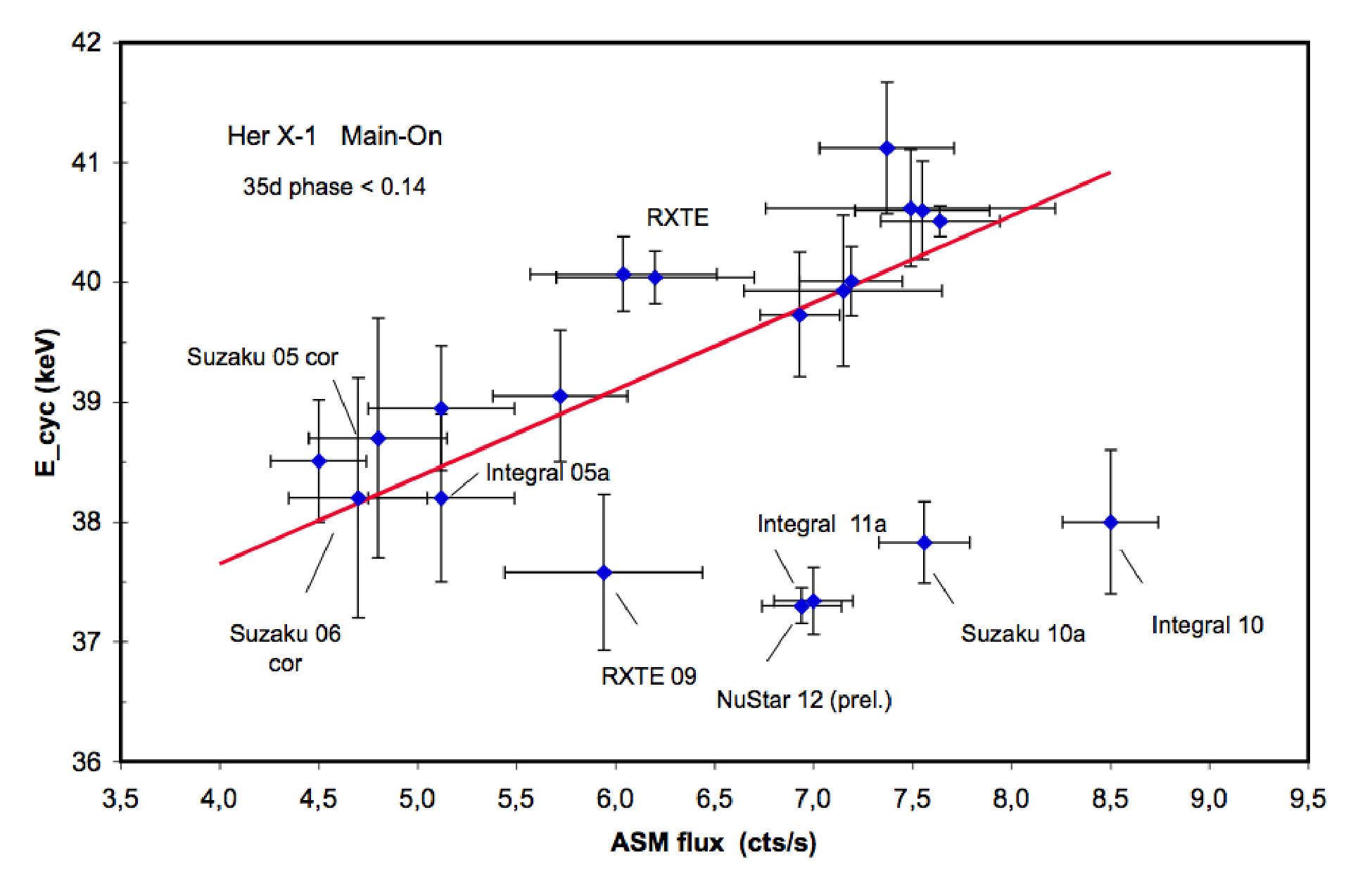}
\hfill
  \includegraphics[width=7.8cm]{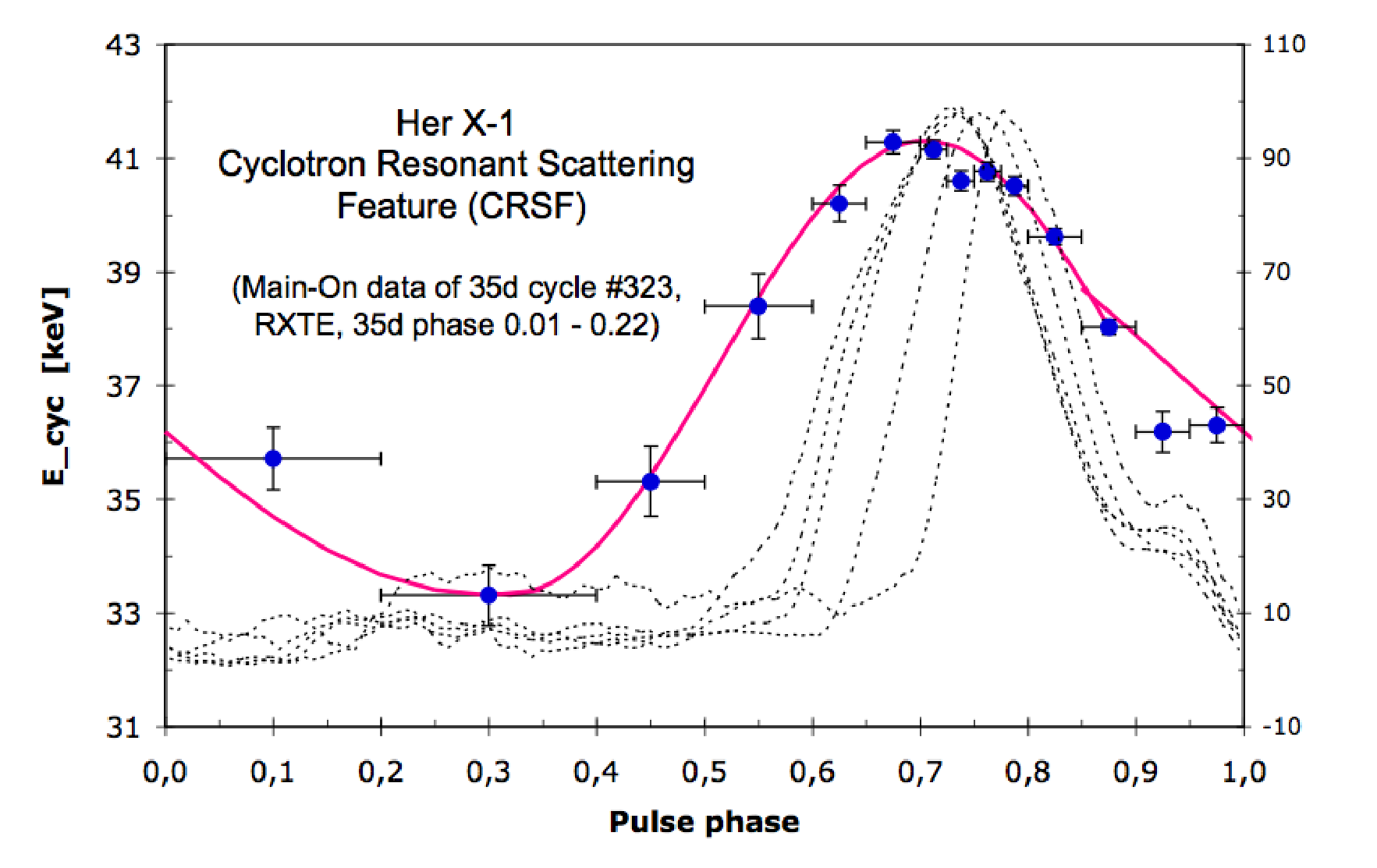}    
%\begin{minipage}[b]{1.0\textwidth}
%\vspace{-10mm}
\caption{\textbf{\textsl{Left}}: The positive correlation between the cyclotron line energy 
  and the maximum X-ray flux of the corresponding 35\,day cycle as  
  measured by \textsl{RXTE/ASM} (see Fig.~2 of Staubert et al., 2007) with eight added points:
  \textsl{INTEGRAL} 2005 (Klochkov et al. 2008), \textsl{Suzaku} of 2005 and 2006 (Enoto et al. 2008), 
  \textsl{RXTE} 2009, \textsl{INTEGRAL} 2010, \textsl{Suzaku} 2010 and \textsl{NuStar} 2012 (preliminary).
  The solid red line is a linear fit, defining a slope of ($0.47\pm0.11$)\,keV/(ASM cts/s). 
  The \textsl{Suzaku} points of 2005/2006 have been 'corrected' upward by 2.8\,keV, 
  to account for the difference arising from the fact that in the analysis by Enoto et al. (2008)
  the Lorentzian profile was used, while for all others the Gaussian profile was used. 
\newline  \textbf{\textsl{Right}}: Mean dependence of cyclotron line energy on pulse phase for the \textsl{Main-On} of 35\,d 
       cycle 323, as observed by \textsl{RXTE}/PCA in 2002 November. The solid red line represents a best fit function
       (a combination of two cosine components). The dotted lines are pulse profiles in the 30-45\,keV range for five
       different 35\,d phases: 0.048, 0.116, 0.166, 0.21 and 0.24. The main pulse is progressively moving to
       the right.}
%\end{minipage}
   \label{correlation}
\end{figure}
%---------------------------------------------------------------------------------------------------------------------

%\vspace{-3mm}
\section{Variation of the cyclotron line energy E$_{\rm cyc}$}

We will discuss the variability in the energy of the Cyclotron Resonant Scattering Feature 
(CRSF) in Her X-1  with regard to the following variables: \\
- Variation with the current X-ray \textbf{luminosity}, both on long and on short time scales \\ 
\indent (the 35\,d precessional period and the 1.24\,s pulse period, respectively). \\
- Variation with \textbf{phase of the 1.24\,s pulsation}. \\
- Variation with \textbf{phase of the 35\,d precessional period}. \\
- Variation with \textbf{time}, that is a true \textbf{secular decay}.
% on a time scales of tens of years.

\subsection{Variation of E$_{\rm cyc}$ with luminosity}

%\vspace{-3mm}
For Her~X-1 the dependence of the centroid energy of the phase averaged cyclotron 
line on X-ray flux was discovered by Staubert et al. (2007) while analyzing an extended
set of observations from \textsl{RXTE}. As reference X-ray flux the maximum of the 
2-10\,keV flux of the respective 35\,d cycle as measured by the All Sky Monitor (ASM) 
on \textsl{RXTE} was used. The correlation was found to be positive, that is the cyclotron 
line energy E$_{\rm cyc}$ increases with increasing X-ray luminosity $L_x$.

The opposite behavior, a negative correlation (a decrease of E$_{\rm cyc}$ with 
increasing $L_x$), had first been noted by Mihara (1995) in observations of a few 
high luminosity transient sources (4U~0115+63, Cep~X-4, and V~0332+53) by \textsl{Ginga}. 
This negative correlation was associated with the high accretion rate during the X-ray 
outbursts, as due to a change in height of the shock (and emission) region above the 
surface of the neutron star with changing mass accretion rate, $\dot{M}$. 
In the model of Burnard et al. (1991), the height of the polar accretion structure 
is tied to $\dot{M}$. From this model one expects that an increase in accretion rate leads 
to an increase in the height of the scattering region above the neutron star surface, and
therefore to a decrease in magnetic field strength and hence E$_{\rm cyc}$. During the 
2004/2005 outburst of V~0332+53 a clear anti-correlation of the line position with X-ray 
flux was observed (Tsygankov et al., 2006).
%, Mowlavi et al. 2006). 
The case of 4U~0115+63, however, is doubtful: while both Nakajima et al. (2006) and 
Tsygankov et al. (2007) had found a general anti-correlation between E$_{\rm cyc}$ and 
luminosity in \textsl{RXTE} data of the Feb--Apr 1999 outburst, M\"uller, S. et al. (2012),
analyzing \textsl{RXTE} and \textsl{INTEGRAL} data from an outburst in March/April
2008, have found that the negative correlation for the fundamental cyclotron line is
likely an artifact due to correlations between continuum and line parameters
when using the NPEX continuum model.

The positive correlation in Her~X-1 was secured by a re-analysis of the same \textsl{RXTE} 
data by Vasco et al. (2011), using the bolometric X-ray flux as reference. This analysis 
confirmed that the originally used 2-10\,keV flux is a good measure of the bolometric
luminosity. While the above discussed analysis tests the correlated variability of
E$_{\rm cyc}$ and $L_x$ on long time scales (35\,d and longer), the "pulse-amplitude resolved"
analysis of Klochkov et al. (2011) does so on short time scales (down to the pulse period
of 1.24\,s). Selecting pulses with amplitudes in certain ranges and producing mean spectra
for each pulse amplitude range, showed that the cyclotron line energy scales positively with
the mean pulse amplitude. In addition, it was found that the photon index $\Gamma$ of the 
underlying power law continuum scales negatively with the pulse amplitude. The same behavior 
was seen in data of the transient A~0535+26. A recent pulse phase resolved analysis of
A~0535+26 observations by \textsl{RXTE} and \textsl{INTEGRAL} showed that data of one 
of the two peaks (of the double peak pulse profile) displays the same trend while data of
the other peak do not (M\"uller, D. et al., 2013).
Applying the same "pulse-amplitude resolved" technique to data of V~0332+53 and 
4U~0115+63, Klochkov et al. (2011) found the same behavior as originally detected in
data sets that were selected on much longer time scales: E$_{\rm cyc}$ decreases and 
$\Gamma$ increases with increasing $L_x$.
Finally, we mention that a positive correlation of E$_{\rm cyc}$ with $L_x$ was also seen by
\textsl{INTEGRAL} in a flare of GX~304-1 (Klochkov et al., 2012).

Our current understanding of the physics behind these correlations assumes that we can
distinguish between \textsl{two accretion regimes} in the accretion column above the polar cap
of the neutron star: \textsl{super- and sub-Eddington accretion}. The former is responsible for
the first detected negative correlation in high luminosity outbursts of transient
X-ray sources: 
%(the reference source being V~0332+53): 
with increasing accretion rate 
$\dot{M}$, the shock and the scattering region move to larger height above the surface of 
the neutron star and consequently to weaker B-field (Burnard et al. 1991). 
Sub-Eddigton accretion, on the other hand, leads to the opposite behavior. In this regime 
an increasing $\dot{M}$ leads to an increased ram pressure on the scattering region and 
a resulting "squeezing" to smaller height and stronger B-field (Staubert et al., 2007).
More detailed physical considerations have recently been presented by Becker et al. (2012).
The persistent source Her~X-1 is clearly a sub-Eddington source.

%Fig.~\ref{correlation} 
Fig.~1(left) reproduces the original correlation graph of Staubert et al. (2007)
with new data points added (see Table~\ref{results}). The E$_{\rm cyc}$ values measured
later than 2007 are significantly lower than the earlier values, while they maintain a similar positive
slope. It is these points that establish the secular decay of E$_{\rm cyc}$ which will be discussed below.

% Position of old Fig. 2

\subsection{Variation of E$_{\rm cyc}$ with pulse phase}
%\vspace{-3mm}
Recently, Vasco et al. (2013) have presented results of pulse phase resolved analysis of 
\textsl{RXTE} observations. 
%Fig.~\ref{Ecyc_pprs} 
Fig.~1(right) shows the mean dependence of the
centroid cyclotron line energy on pulse phase for all \textsl{Main-On} data of cycle 323
(for 35\,d phases between 0.1 and 0.22). It is the first time that for Her~X-1 a phase profile 
of the centroid cyclotron line energy E$_{\rm cyc}$ is produced with such a high resolution 
and statistical accuracy (the four smallest bins around the peak of the pulse have a width 
of 1/80 of a phase). The shape of the modulation is nearly sinusoidal, but for a good fit,  
at least two sine functions are needed (solid red line). The peak to peak amplitude of the 
modulation is $\sim21$\% of the mean energy. There is no 35\,d dependence of the 
shape of this modulation. Generally, the E$_{\rm cyc}$ profile roughly follows 
the shape of the pulse profile, with its broad maximum close to the maximum 
of the main peak (pulse phases 0.7-0.8). The dashed curves are 30-45\,keV pulse profiles 
for five different 35\,d phases: the main peak is progressively moving to the right with
increasing 35\,d phases.\footnote{The definition of pulse phase zero is tied to the "sharp edge"
at the decay of the right shoulder of the main peak (see Staubert et al., 2009b, 2013).} 
The fact that the pulse moves, while the shape of the E$_{\rm cyc}$ profile stays constant,
leads to the expectation for a variation in pulse phase averaged E$_{\rm cyc}$ with 35\,d
phase (see Section 3.3).

Pulse phase modulation in E$_{\rm cyc}$ is quite  common among accreting X-ray pulsars 
and is generally believed to be due to the changing viewing angle under which the X-ray 
emitting regions are seen. If we adopt the idea, that peaks in the pulse profile can be 
associated with beamed radiation emitted from the accreting regions at the magnetic 
poles, then the observed pulse profiles are a representation of the emission characteristics. 
During a full neutron star rotation we may be seeing the beamed radiation of different 
emitting spots under changing angles. In addition, variable absorption and reflection at the
inner edge of the precessing accretion disk and possibly a variable orientation due to 
precession of the neutron star (Postnov 2004) may contribute. The situation can be further 
complicated by gravitational bending. It is therefore not straightforward to interpret the 
observed pulse profiles  and the variation in E$_{\rm cyc}$ in terms of accretion 
geometry and beaming characteristics. We note here, that also the photon index 
$\Gamma$ of the power law continuum is dependent on pulse phase with a distinct 
minimum close to the pulse peak (Vasco et al., 2013).

% Position of of Fig. 3

\subsection{Variation of E$_{\rm cyc}$ with precessional phase}
%\vspace{-3mm}

A few measured low E$_{\rm cyc}$ values at late 35\,d phases had led us to suspect that the
pulse phase averaged cyclotron line energy may not be constant over the 35\,d cycle. This has triggered
proposals for repeated observations of Her~X-1 at different 35\,d phases by \textsl{INTEGRAL}, \textsl{Suzaku}
and most recently \textsl{NuStar}. These observations are not yet fully analyzed, but using data from earlier
\textsl{RXTE} observations (cycle 323), we have already clear indications, that there is a weak modulation
of E$_{\rm cyc}$ over the \textsl{Main-On}. Vasco et al. (2013) had shown, that the shape of the 
(E$_{\rm cyc}$ vs pulse phase)-profile is not dependent on 35\,d phase, while the pulse profile itself is.
%Fig. \ref{Ecyc_pprs} 
Fig.~1(right) demonstrates that the 30-45\,keV pulse profiles (this is the energy range which includes
most of the CRSF) clearly move to the right with increasing 35\,d phase, which means that more and more
photons are found at pulse phases where the cyclotron line energy is decreasing. So, a modulation of
E$_{\rm cyc}$ with 35\,d phase is inevitable. In order to quantitatively test this, we have performed a formal 
folding of 30-45\,keV pulse profiles (for 10 different 35\,d phases) with template 
(E$_{\rm cyc}$ vs pulse phase)-profiles at the same 35\,d phases. These template profiles were constructed
by inter- and extrapolation of the four individual profiles at different 35\,d phases given in Fig. 3 and 4 of 
Vasco et al. (2013), taking into account that both the maximum value and the peak-to-peak amplitude are 
slightly 35\,d phase dependent.
By folding the pulse profile with the corresponding E$_{\rm cyc}$ profile, the expected pulse phase averaged 
E$_{\rm cyc}$ value can be calculated. 
%In Fig.~\ref{Ecyc_folding} 
In Fig.~2(left) these calculated values are shown as blue triangles
(connected by the solid blue line): a slow increase up to phase $\sim0.19$ is followed by a somewhat sharper decay.
For comparison, we show the directly measured phase averaged E$_{\rm cyc}$ values (data points with uncertainties)
for 13 small integration intervals covering the \textsl{Main-On} of cycle 323. The directly measured values are
consistent with the calculated modulation. With the final analysis of additional data, more definite conclusions are
expected. This new result adds another piece to the puzzle on the question about the
physical nature of the 35\,d modulation - precession of the accretion disk (and?) precession of the neutron star? -
and the mechanism of generating the varying pulse profiles and the varying spectra - CRSF and continua
(see discussion in Staubert et al., 2009a, 2013 and Vasco et al., 2013).
%Even though, we would not claim a modulation based on the directly measured values alone

%----------------------------------------------------------------------------------------------------------------
% Fig. 4
\begin{figure}
\includegraphics[width=7.8cm]{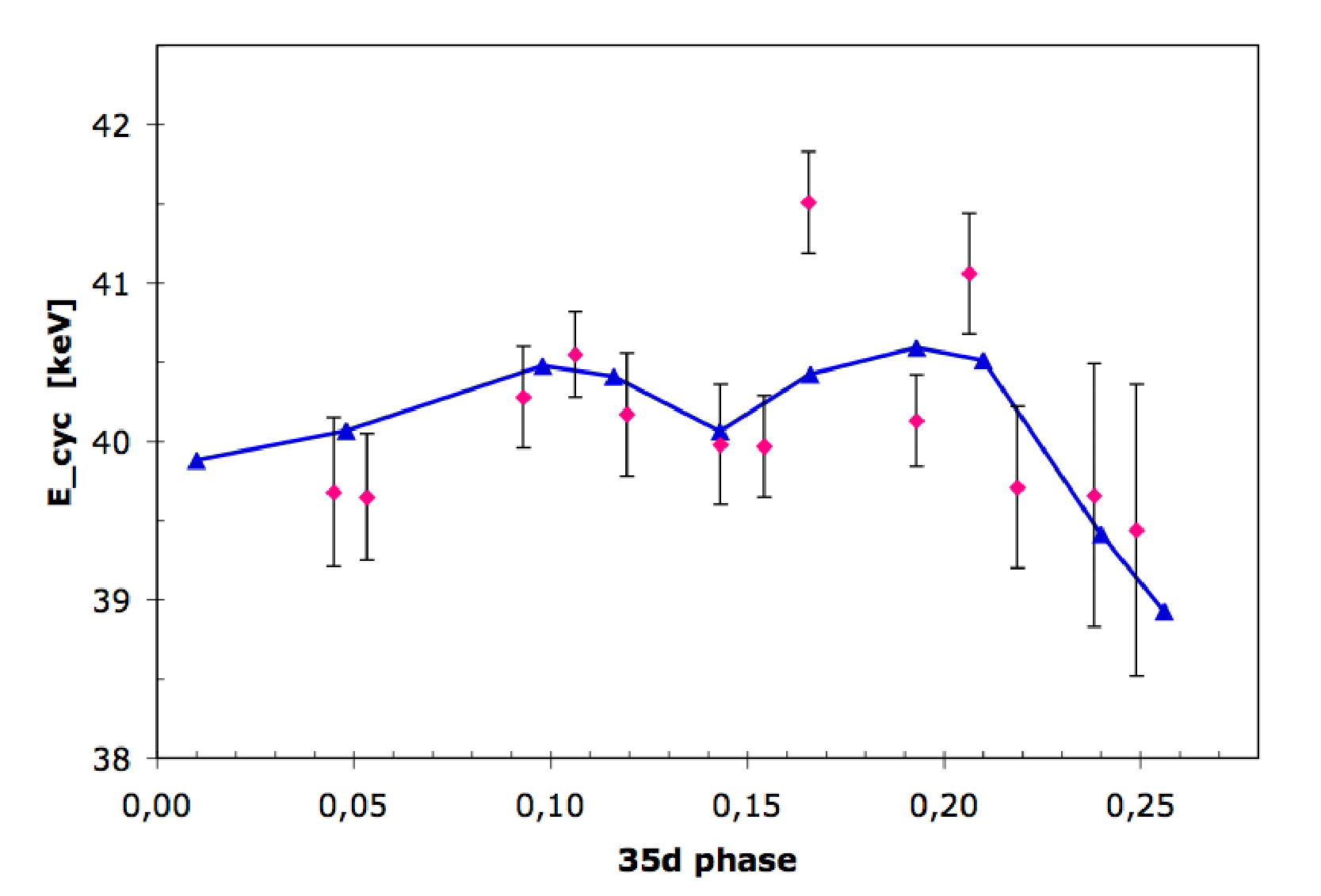} 
\hfill
\includegraphics[width=7.8cm]{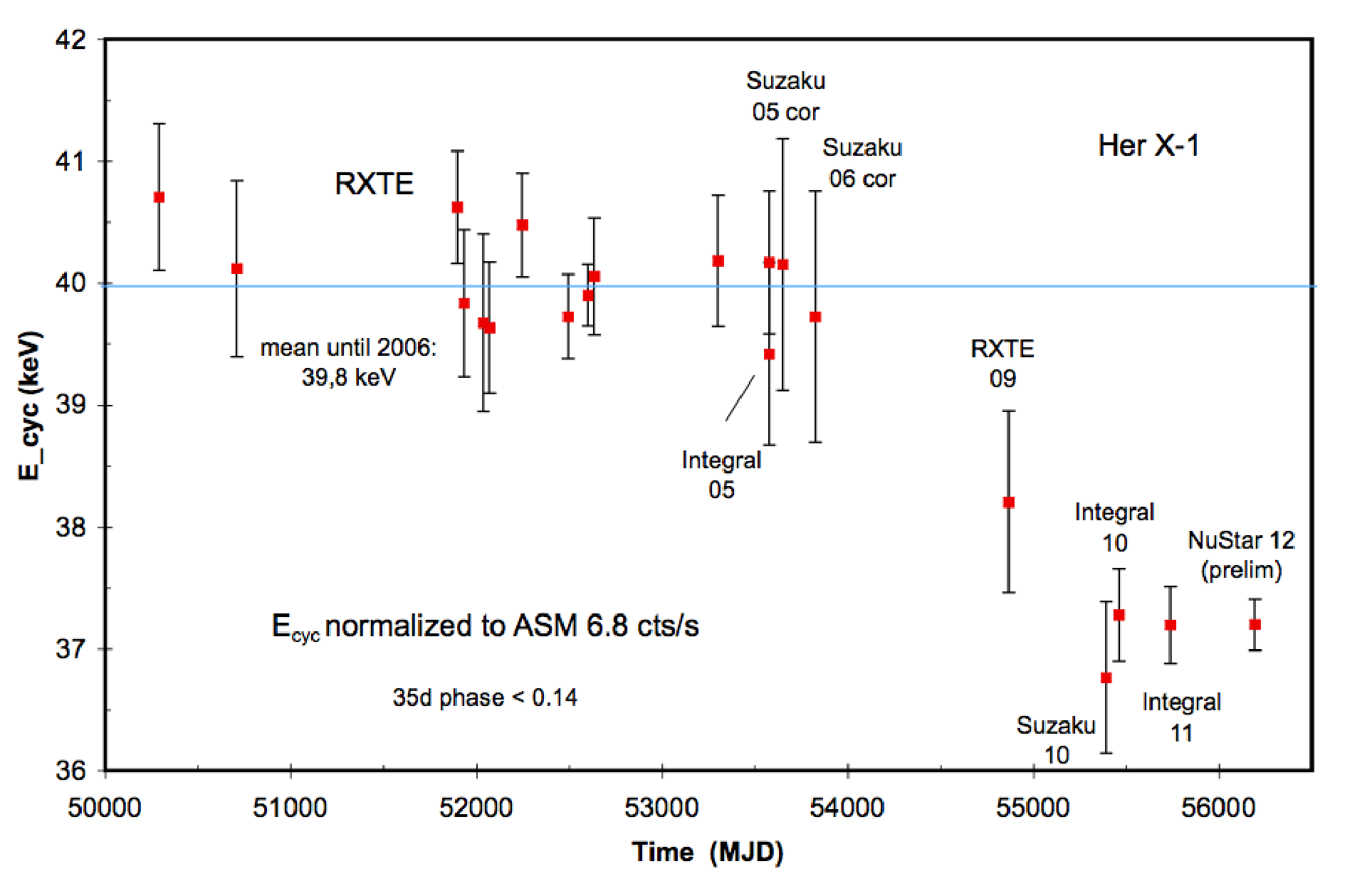}
\hfill
%\begin{minipage}[b]{1.0\textwidth}
\vspace{-5mm}
\caption{\textbf{\textsl{Left}}: Centroid pulse phase averaged cyclotron line energies at different 35\,d phases of 
    \textsl{Main-On} cycle 323. The data points with uncertainties are direct measurements for ten small integration 
    intervals. The blue triangles connected by the solid blue line are values which are calculated by folding observed 
    pulse profiles in the 30-45\,keV range with template (E$_{\rm cyc}$ vs pulse phase)-profiles for the same 35\,d phases.
\newline \textbf{\textsl{Right}}: Her~X-1 pulse phase averaged cyclotron line energies E$_{\rm cyc}$, normalized
    to a reference ASM count rate of 6.9\,cts/s using the originally found correlation function.
    A break in E$_{\rm cyc}$ is apparent around MJD 54000. Note, that the four points measured last are all low, 
    indicating a true secular variation.}
%\end{minipage}
   \label{normalized}
\end{figure}
%----------------------------------------------------------------------------------------------------------------

\subsection{Variation of E$_{\rm cyc}$ with time - secular variation}
%\vspace{-3mm}
In investigating the long-term behavior of the pulse phase averaged centroid
cyclotron line energy, Staubert et al. (2007) had discovered that E$_{\rm cyc}$
depends on the X-ray flux of the source. Their Fig.~1 compiles the historical
measurements until this time: after the initial measurements in the range 34--37\,keV, 
a (so far unexplained) upward jump occurred, between 1991 and 1994, to beyond 
42\,keV, followed by an apparent decrease with time. This decrease had been noticed
after the first observations with \textsl{RXTE} in 1996 and 1997 (in comparison to
values from \textsl{CGRO}/BATSE and \textsl{Beppo}/SAX), and had served as an
important argument to ask for more observations of Her~X-1. 
%(which quite fortunately were granted by the Time Allocation Committees). 
But then, Staubert et al. (2007) showed that this decrease with time was largely 
an artifact due to the dependence of E$_{\rm cyc}$ on X-ray flux. 
Plotting $E_{\rm cyc}$ against the 2-10\,keV X-ray flux, as measured by ASM on 
\textsl{RXTE} shows a clear positive correlation, with an increase in E$_{\rm cyc}$ 
by $\sim 7$\% for a factor of two increase in flux (see Section~3.1).
So, here nature had conspired such that later measurements were (on average) 
taken when the flux happened to be low (Her X-1 is known for varying
its flux within a factor of two, on time scales of a few 35\,d cycles).

Now, however, the situation has changed. Here, we report about
the first statistically significant evidence for a \textsl{true secular decay of the
phase averaged cyclotron line energy}. From 
%Fig.~\ref{correlation}, 
Fig.~1(left), it is already evident that E$_{\rm cyc}$ values measured after 2008 are significantly lower,
while maintaining a similar positive correlation within themselves (possibly less steep). 
In order to separate the dependence on time and the dependence on X-ray flux, 
we have normalized the measured E$_{\rm cyc}$ values to a common X-ray flux 
(namely 6.8 ASM cts/s), using the slope as found in the original flux correlation. 
The result is shown in Fig~2(right).
%Fig.~\ref{normalized}. 
The normalized E$_{\rm cyc}$ values until 2006 are consistent with a constant value of 39.8\, keV, 
a significant decrease occurs after that. This decrease is consistent with a linear decay with time between 
2006 and 2012, but a more abrupt change can not be excluded. We note, that all 
measurements used for this plot are from observations corresponding to the 
\textsl{Main-On} of Her X-1 at 35\,d phases less than 0.14 (for which the dependence 
on 35\,d phase is very weak).

With regard to the physical implications of the now observed secular decrease of the 
cyclotron line energy, we speculate that the phenomenon could be connected to a 
%long-term
restructuring of the polar magnetic field of the neutron star. This may be 
due to an adjustment of the field when coping with the continued accretion of plasma.
Theoretically, such scenarios have been investigated e.g., by Brown and Bildsten (1998), 
Litwin et al. (2001), Mukherjee and Bhattacharya (2012) and  Mukherjee et al. (2012).

%\begin{thebibliography}{99}
%\bibitem{...} 
%\end{thebibliography}

%\vspace{15mm}

%\bibliographystyle{aa}
%\vspace{-3mm}
%\bibliography{refs_herx1}

\end{document}